\documentclass[amsmath,amssymb,aps,twocolumn,prb,longbibliography,floatfix,superscriptaddress]{revtex4-2}
\usepackage{graphicx,bm,times}
\usepackage[utf8]{inputenc}
\usepackage{braket}
\usepackage{amsfonts}
\usepackage{color}

\usepackage[breaklinks]{hyperref}
\hypersetup{
    pdfstartview={FitH},
    colorlinks=true,
    linkcolor=blue,
    citecolor=blue,
    urlcolor=blue
}

\begin{document}

\title{Exposing Altermagnetism through Momentum Density Spectroscopy}

\author{Wenhan~Chen}
\affiliation{Max-Born-Institut für Nichtlineare Optik und Kurzzeitspektroskopie, Max-Born-Strasse 2A,
12489 Berlin, Germany}
\affiliation{Max Planck Institute of Microstructure Physics, Weinberg 2, D-06120 Halle, Germany}
\affiliation{H.~H.~Wills Physics Laboratory, University of Bristol, Tyndall Avenue, Bristol, BS8 1TL, United Kingdom}

\author{Alyn~D.~N.~James}
\affiliation{H.~H.~Wills Physics Laboratory,
University of Bristol, Tyndall Avenue, Bristol, BS8 1TL, United Kingdom}

\author{Stephen~B.~Dugdale}
\affiliation{H.~H.~Wills Physics Laboratory, University of Bristol, Tyndall Avenue, Bristol, BS8 1TL, United Kingdom}

\date{\today}

\begin{abstract}
Materials which show a strong time-reversal symmetry-breaking response leading to spin-polarization phenomena, in conjunction with antiparallel magnetic alignments producing zero net magnetization, have recently been identified, classified, and been given the name `altermagnets'. However, measuring and diagnosing possible candidates as altermagnetics still remains a challenge. From the uncertainty of the material being an altermagnet, additional experimental probes are essential to resolve this. Here, we propose using spin-dependent and magnetic momentum density probes such as  spin-polarised positron annihilation and revisiting magnetic Compton scattering. By looking at the previously claimed altermagnetic candidates RuO$_2$, CrSb and MnTe, we present theoretical altermagnetic calculations of the experimental quantities measured by these probes. We show that these quantities should produce a measurable signal and unequivocally confirm the altermagnetic state. We also highlight the additional benefits from these probes such as extracting spin-resolved Fermi surfaces which are key for further understanding the nature of the altermagnetic state.

\end{abstract}

\maketitle

\section{Introduction}

The discovery of materials whose magnetism has some characteristics more typically associated with ferromagnets, and other properties usually associated with antiferromagnets, has forced a reconsideration of this historical dichotomy \cite{PhysRevX.12.040501,PPR:PPR404397}.
Ferromagnets possess a spin-polarization in reciprocal (momentum) space associated with the magnetization in real space and this produces electronic band structures with broken time-reversal symmetry and spin-split energy bands. As pointed out by \v{S}mejkal {\it et al.}\cite{PhysRevX.12.040501}, in many ways antiferromagnets, whose compensated antiparallel magnetic moment alignment leads to a net zero magnetization, are like nonmagnetic materials in terms of their visibility to macroscopic or electric probes commonly used to investigate ferromagnets, such as anomalous hall effect measurements~\cite{RevModPhys.82.1539,vsmejkal2022anomalous,liu2025different}. Until recently \cite{PhysRevX.12.031042}, collinear magnets were classified as either ferromagnets (displaying a net magnetic moment) or antiferromagnets.
Motivated by a wide body of theoretical work (see Ref.~\cite{PhysRevX.12.040501} and references therein) which predicted time-reversal symmetry breaking and spin-split band structures (typically characteristic of ferromagnets) in materials with zero net magnetic moment (characteristic of antiferromagnets), \v{S}mejkal et al. \cite{PhysRevX.12.040501} made a detailed classification of magnetic phases based on a generalized symmetry formalism. This classification established a hitherto unknown magnetic phase to sit alongside ferromagnets and antiferromagnets. 
Coined `altermagnetism', this magnetism is characterized by sublattices with opposite spin connected by a crystallographic rotational symmetry which give rise to a net zero magnetic moment 
\cite{PhysRevX.12.040501,PhysRevB.106.094432} in real space. In reciprocal (momentum) space, however, the same rotational symmetries lead to an unconventional order in the spin-polarization and a unique spin-momentum locked phase~\cite{PPR:PPR404397}. Of particular significance to the present study is the fact that there is an element of the point group which does not map the Fermi surface for a given spin onto itself, but rather onto the Fermi surface of the opposite spin \cite{mazin2022notes}. The net-zero magnetic moment and unique spin-momentum-locked bands (and Fermi surface) properties make it difficult to distinguish altermagnetism from ordinary antiferromagnism using conventional transport and spectroscopic measurements. This poses a significant challenge for the unambiguous identification (or `fingerprinting') of altermagnetism. Experimentally, inelastic neutron scattering and spin-resolved ARPES have each been shown to probe the characteristic magnon fingerprints~\cite{PhysRevLett.133.156702,PhysRevLett.134.226702} and the distinctive band splittings in parts of the Brillouin zone~\cite{osumi2024observation,reimers2024direct} that arise from altermagnetic spin splitting. However, these results have not unequivocally proved the altermagnetic state in proposed candidates. Theoretically, quantum oscillations have also been proposed as another fingerprinting approach, although practical application remains limited by sample quality and intrinsic material properties~\cite{li2025diagnosing}. 

Spectroscopies such as Compton scattering and positron annihilation which probe the \textit{bulk} electronic structure of materials through their electron momentum density (EMD) \cite{dugdale_2014} are ideally suited to revealing the presence of altermagnetism. In a Compton scattering experiment, the ground-state momentum distribution is extracted from the energy distribution of X-ray photons which have been inelastically scattered in the sample being studied. The so-called Compton profile is a one-dimensional projection (integration over two momentum components) of the EMD, ($\rho(\mathbf{p})$, where $\mathbf{p}$ is the real momentum) onto whatever crystallographic direction is aligned with the scattering vector. In a positron annihilation experiment, positrons from a radioisotope source are implanted into the sample being studied and, after rapidly coming into thermal equilibrium, propagate in their own delocalised Bloch states before annihilating with an electron. The annihilation process predominantly produces two $\gamma$-photons which are emitted in almost anti-parallel directions. By measuring the distribution of angular deviations of the photons from being anti-parallel (using a pair of position-sensitive detectors (e.g. \cite{Dugdale_2013}), two components of the momentum of the electron-positron pair can be measured. The momentum is completely dominated by the electron's momentum and thus in a two-dimensional angular correlation of annihilation radiation (2D-ACAR) experiment, it is a two-dimensional projection (integration over one momentum component) of the momentum density which is measured.  However, to emphasise that this is the electron momentum distribution {\it as seen by the positron}, the terms `electron-positron momentum density' or (more commonly) the `two-photon momentum density' (TPMD) are used. 

There are special versions of both spectroscopic techniques which are sensitive to magnetism. Magnetic Compton scattering measures spin-density in momentum space (difference between the momentum densities for spin-up and spin-down electrons) by taking advantage of a spin-dependent term in the scattering cross-section for photons with a component of circular polarization \cite{Duffy_2013}. This contribution is usually isolated by making measurements in which the magnetic field is reversed with respect to the scattering vector. In the case of positron annihilation, spin-polarized 2D-ACAR exploits a consequence of parity violating weak-decay which births the positron and which gives it a spin-polarization preferentially parallel to its velocity vector. Since two-photon annihilation involves an electron with spin opposite to that of the positron, the positron will preferentially annihilate with one of the two spin populations. As explained later, this permits, within certain approximations, the momentum distributions of spin-up and spin-down electrons to be isolated \cite{PhysRevB.34.5009, PhysRevB.42.1533, Genoud_1991,PhysRevLett.115.206404}. 
Signatures of the Fermi surface appear as discontinuities in the momentum density which occur when an energy band crosses the Fermi energy and ceases to contribute to $\rho(\mathbf{p})$. Sensitivity to the Fermi surface is particularly evident in positron annihilation measurements, since the positively charged positron is strongly screened by the valence electrons and preferentially annihilates with an electron at the Fermi surface. Therefore, positron annihilation could unambiguously demonstrate the presence of altermagnetism by revealing this distinguishing feature of the spin-resolved Fermi surface. 
The spin-resolved positron annihilation spectra are not limited to certain factors such as low temperature, high magnetic fields and the dependence of the Liftshitz transition as with the quantum oscillations proposed by Ref.~\cite{li2025diagnosing}. Indeed, spin-resolved positron annihilation directly observes the Fermi surfaces in $p$-space (and thus $k$-space). Furthermore, we show that this technique is capable to observe the altermagnetic state in insulating materials.

The idea of using the electron momentum distribution to reveal altermagnetism is not new. Indeed, Bhowal and Spaldin \cite{PhysRevX.14.011019} have championed the potential of the electron momentum density as a route to revealing the presence of altermagnetism. Proposing that these altermagnetic materials exhibiting unconventional antiferromagnetism (which they point out can be conveniently described in terms of ferroic ordering of magnetic octupoles), they suggested that the measurement of particular oscillatory signatures in magnetic Compton profiles in the classic two-sublattice antiferromagnet MnF$_2$ could be used for experimental verification, although their predictions indicate that the magnetic Compton profiles of this material may be too small to be measurable (and this material was subsequently shown not to be an altermagnet ~\cite{PhysRevLett.134.226702}). This followed on from a study in which they showed that in the absence of inversion symmetry there would be a non-zero antisymmetric magnetic Compton profile even in non-magnetic materials \cite{PhysRevLett.128.116402}. Using ferroelectric PbTiO$_3$ as an example, they showed how the electric dipole in real space is equivalent to a magnetoelectric toroidal moment in reciprocal space, giving rise to antisymmetric magnetic Compton profiles (with zero integral, since there is no net magnetic moment) and which changes sign on reversal of the electric field. 

In this paper we present a proposal for how altermagnetism can be clearly identified through measurements of the spin-resolved TPMDs and magnetic Compton profiles (MCPs) of three putative altermagnetic candidates: RuO$_2$, CrSb and MnTe~\cite{Bai_AM_review}. The altermagnetic state of these materials have claimed experimental evidence ~\cite{lin2024observation,OF-RuO2-ARPES-AM,reimers2024direct,yang2025three,PhysRevLett.132.176701,lee2024broken,osumi2024observation}, although these claims for RuO$_2$ have been challenged by the interpretation of other experimental measurements ~\cite{kessler2024absence,liu2024absence,smolyanyuk2025origin}. The purpose of our study is to provide momentum density signatures of the altermagnetic state, not to claim whether the materials calculated here are altermagnetic in nature or not. 
We emphasise the use magnetic Compton scattering and a positron annihilation experiments to unambiguously identify the altermagnetic state of candidate materials. Further unique information of the electronic and magnetic can be extracted from these measurements such as observing the $\textit{spin-resolved}$ Fermi surfaces.

\section{Background}

The EMD, $\rho_\sigma(\mathbf{p})$, is formally derived from the Fourier transformed real-space wave functions $\psi^\sigma_{\mathbf{k}, \eta}(\mathbf{r})$ : 
\begin{equation}
  \rho_\sigma(\mathbf{p})= \sum_{\mathbf{k}, \eta}n^{\sigma}_{\mathbf{k}, \eta}\left|\int\limits_{V} \exp (-i \mathbf{p} \cdot \mathbf{r}) \psi^\sigma_{\mathbf{k}, \eta}(\mathbf{r}) \mathrm{d} \mathbf{r}\right|^{2}, \label{EMD}
\end{equation} 
where $n^{\sigma}_{\mathbf{k}, \eta}$ are occupation distribution with eigenstate (band) index $\eta$, spin $\sigma$. The electron momentum density within the DFT formalism is computed using the Kohn-Sham wave functions and occupation functions.  

The Fermi surface information encoded in $\rho_{\sigma}(\mathbf{p})$ is distributed throughout real momentum- ($p$-)space, but the discrete translational invariance of crystal momentum ($k$-) space can be restored by `folding' the $p$-space distribution back into the first Brillouin zone. This is done by translating contributions at momenta outside the first Brillouin by the appropriate reciprocal lattice vector to bring them back into the first Brillouin zone. This process, often referred to Lock-Crisp-West or LCW folding~\cite{Lock_1973}, is similar to moving from the extended zone scheme to the reduced zone scheme. The resulting distribution is simply the occupation distribution $n(k)$ or, in the case of positron annihilation, the occupation distribution {\it seen by the positron} $n^{2\gamma}(k)$. The Fermi surfaces (which are just the {\it loci} of the step-like changes in the occupation number) of a large number of materials have been determined using these distributions \cite{PhysRevLett.79.941,PhysRevLett.96.046406,Fretwell_1995,PhysRevLett.124.046402}.

The Compton profile $J(p_{z})$ (i.e., the doubly-projected EMD) is measured or calculated along the scattering vector (which is conventionally labelled $p_z$) :
\begin{equation}
J(p_{z})= \sum_{\sigma} \iint \rho_{\sigma}(\mathbf{p}) \mathrm{d} p_{x} \mathrm{d} p_{y}. \label{eq:J}
\end{equation}
In analogy to the Compton profile, the magnetic Compton profile (MCP) $J_{\mathrm{mag}}(p_z)$ is defined as the 1D projection of the spin-polarized electron momentum density:
\begin{equation}
 \label{eq:J_mag}
  J_{\mathrm{mag}}(p_z)= \iint 
    \left[
      \rho_\uparrow(\mathbf{p})-\rho_\downarrow(\mathbf{p})
    \right]
    \mathrm{d}p_x \mathrm{d}p_y\,.
\end{equation} 

Switching our discussion to positron annihilation, the connection between the (real space) wave functions and the TPMD can be simply understood via the Fourier transformation~\cite{PhysRevB.71.233103} (assuming spin degeneracy for clarity) as 
\begin{equation}
\begin{aligned}
  \rho^{2\gamma}(\mathbf{p}) &= \sum_{\mathbf{k}, \eta}n_{\mathbf{k}, \eta}\left|\int\limits_{V} \sqrt{\gamma(\mathbf{r})}\psi_{\mathbf{k}, \eta}^{e}(\mathbf{r})\psi^{p}(\mathbf{r}) \exp (-i \mathbf{p} \cdot \mathbf{r})  \mathrm{d} \mathbf{r}\right|^{2}\\
  &= \sum_{\eta}n_{\mathbf{k}, \eta}\left|C_{\mathbf{k}+\mathbf{G},\eta}\right|^{2},
\end{aligned}
\label{TPMD}
\end{equation}
where $\psi^{p}(\mathbf{r})$ is the positron wave function and $\psi_{\mathbf{k}, \eta}^{e}(\mathbf{r})$ is the Bloch electron wave function with eigenstate index $\eta$ at point $\mathbf{k}$ in the $k$-mesh used. The quantity $\gamma(\mathbf{r})$ is the so-called electron-positron enhancement factor which is designed to describe the electron-positron correlations \cite{PhysRevB.82.125127}. If this factor is assumed to be a constant, then this is referred to as the independent particle model. However, different approximations can be adopted to improve agreement between experiment, such as the Drummond enhancement~\cite{drummond_2010,drummond_2011} which was used in this study. The last line of Eq.~\ref{TPMD} introduces $C_{\mathbf{k}+\mathbf{G},\eta}$ which is a shorthand for the Fourier coefficients of the combined enhanced electron-positron wave function product.

A 2D-ACAR experiment with an unpolarized positron will measure the TPMD integrated over one component of momentum defined by the detector-sample-detector axis of the spectrometer (usually defined as `$z$' axis) :
\begin{equation}
\begin{aligned}
  M(p_x,p_y) \propto \sum_{\sigma} \int_{-\infty} ^\infty \rho^{2\gamma}(\mathbf{p}) dp_z.
\end{aligned}
\label{2D-ACAR}
\end{equation}
Since two-photon annihilation involves an electron annihilating with a positron of opposite spin, if the positron source has a polarization and the electron bands are not degenerate, measurements made with a magnetic field parallel and anti-parallel to the positron emission direction will be different from each other. Defining the fraction of polarized positrons, $F$, as being the fraction parallel to the sample magnetization, a typical radioisotope positron source such as $^{22}$Na has an $F$ of $\sim 0.7$. 
Following closely the approach described in Ref.~\cite{PhysRevLett.115.206404}, separate 2D-ACAR measurements can be made with the magnetic field parallel ($p$) and antiparallel ($a$) to the direction in which the positrons are emitted from the source. Making the very reasonable assumption that $3\gamma$ annihilations can be neglected \cite{PhysRevB.42.1533}, this will now yield spectra $M^{p/a} (p_x,p_y)$ of the form :
\begin{equation}
\begin{aligned}
  M^{p/a} (p_x,p_y) \propto F \frac{\rho^{2D}_{\uparrow/\downarrow}(p_x,p_y)}{\lambda_{\uparrow/\downarrow}} + (1-F)\frac{\rho^{2D}_{\downarrow/\uparrow}(p_x,p_y)}{\lambda_{\downarrow/\uparrow}},
\end{aligned}
\label{magnetic-2DACAR}
\end{equation}
where $\rho^{2D}_{\uparrow/\downarrow}(p_x,p_y)$ are, respectively, the once-integrated spin-up and spin-down momentum densities and $\lambda_{\uparrow/\downarrow}$ are respectively the annihilation rates of the positron with spin-up and spin-down electrons. 
It follows that if both $M^p$ and $M^a$ are measured, and $F$ has been independently determined (see, for example \cite{ceeh_dissertation}), then simple algebra shows that both $\rho^{2D}_{\uparrow}$ and $\rho^{2D}_{\downarrow}$ can be isolated :

\begin{equation}
\begin{aligned}
  \rho^{2D}_{\uparrow}(p_x,p_y) \propto F M^{p} - (1-F) M^{a},\\
  \rho^{2D}_{\downarrow}(p_x,p_y) \propto (F-1) M^{p} + FM^{a}.
\end{aligned}
\label{magnetic-2DACAR-isolated}
\end{equation}

If the spin-up and spin-down projected momentum distributions can be obtained, insight into altermagnetic state can be obtained by first examining the spin-resolved so-called `radial anisotropy', which is just the difference between the projected (2D) momentum density $\rho^{\rm 2D}(p_{x},p_{y})$ and its average value evaluated on concentric circles. This highlights the deviation from the angular average at constant momentum. 
The radial anisotropy of $\rho^{\rm 2D}(p_{x},p_{y})$ is given by (assuming spin degeneracy for clarity),
\begin{equation}
R(p_{x},p_{y})=\rho^{\rm 2D}(p_{x},p_{y})-\overline{\rho^{\rm 2D}(p_{x},p_{y})}|_{p=\text{const.}},
\end{equation}
where $\overline{\rho^{\rm 2D}(p_{x},p_{y})}|_{p=\text{const.}}$ is the isotropic distribution constructed from the angular average of $\rho^{\rm 2D}(p_{x},p_{y})$.
The radial anisotropy of $\rho_(p_{x},p_{y})$ contains contributions from both the fully and any partially occupied bands. In a metal, some features will be due to the presence of the Fermi surface. 

A projection of the spin-resolved occupation number (in $k$-space) in the first Brillouin zone could be obtained by applying the LCW folding procedure~\cite{Lock_1973} to the spin-resolved momentum distributions projections. Thus for extracting the spin-resolved Fermi surfaces, the ideal experiment would comprise a set of similar measurements made with their projections along different crystallographic direction, so that the full 3D momentum density could be obtained by tomographic reconstruction (see, e.g. \cite{Kontrym-Sznajd1990}) and thus the Fermi surfaces of the spin-up and spin-down electrons are isolated, as previously done by Weber {\it et al.} \cite{PhysRevLett.115.206404}.

For completeness, we highlight some details about the experimental techniques which would have to be considered for measuring altermagnetic candidates. The sample would need to be a single crystal with dimensions ideally of the order of a few millimetres for Compton scattering and 2D ACAR; 
this is required to deliver an appreciable count rate. 
This does focus attention on the typical domain sizes of altermagnetic materials, which can be on the scale of microns as in MnTe~\cite{Rikako_2025_AM_MnTe}. However, other experimental techniques are able to distinguish an altermagnetic signal when measuring multiple domains~\cite{krempasky2024altermagnetic,hariki2024x}, so similar sample preparation could be used for Compton scattering and 2D ACAR. Finally, the high sample quality is required for 2D ACAR to help nullify the unwanted signal of positron annihilation from defects such as site vacancies~\cite{Dugdale_2014_ACAR_defects}.

\section{Methods}

All the electronic structure calculations were performed using the full potential APW+lo {\sc elk} DFT code~\cite{elk}. Each material calculation used the Perdew-Burke-Ernzerhof (PBE) generalized gradient approximation (GGA) functional~\cite{PhysRevLett.77.3865}. In order to obtain the altermagnetic state, the symmetry of the Ru, Cr and Mn atomic positions was broken in their respective calculations, so that they were treated as being inequivalent~\cite{sym_split,PhysRevLett.118.077201}. 
Spin-orbit coupling (SOC) is not included in any of the calculations as it has been shown that it does not significantly change the altermagnetic spin texture~\cite{guo2023spin}. We note that we only focus on the signatures of the altermagentic state within the momentum densities which will be displayed in our calculations. The finer details may be adjusted with the inclusion of SOC or other factors.

To obtain the metallic altermagnetic electronic structure of tetragonal RuO$_2$ with lattice parameters $a=4.492 \rm \AA$ and $c=3.1061 \rm \AA$~\cite{Kyo-Hoon_2019,PhysRevLett.118.077201}, the calculation was done within the DFT+U framework~\cite{Liechtenstein_1995,Shick_1999,Petukhov_2003,Bultmark_2009}. This DFT+U calculation used the Hubbard Hamiltonian term parameterised with $U=2.8$~eV, $J=0.2$~eV and the around-mean-field (AMF) approximation to the double counting term. These were the parameters for the RuO$_2$ calculation presented in Ref.~\cite{Kyo-Hoon_2019}. The DFT+U RuO$_2$ calculation was converged on a $ 20 \times 20 \times 30 $ Monkhorst-Pack $k$-mesh of 3330 irreducible $k$-points in the first Brillouin zone.

The metallic altermagetic electronic structure of hexagonal CrSb (space group $P6_3/mmc$) \cite{Willis:a00902} with lattice parameters $a= 4.121 \rm \AA$ and $c= 5.467 \rm \AA$~\cite{radhakrishna1996inelastic,PhysRev.129.2008} was obtained within the DFT framework. 
The calculation was converged on a $20 \times 20 \times 15 $ Monkhorst-Pack $k$-mesh of 1155 irreducible $k$-points in the first Brillouin zone. 

For the hexagonal MnTe (space group $P6_3/mmc$) with lattice parameters $a=4.158 \rm \AA$ and $c=6.726 \rm \AA$ ~\cite{li2022AFM_MnTe}, the same $k$-mesh as the CrSb calculation was used. The DFT+U framework was used to obtain an insulating electronic structure for the purpose to emphasise that the ACAR and magnetic Compton scattering measurements are sensitive to the altermagnetic state within insulators too. 
Indeed, previous studies showed that MnTe is a semi-conductor~\cite{PhysRevB.61.13679,PhysRevMaterials.3.025403}. 
It should be noted that the altermagnetic state is present within the DFT(GGA) calculation of MnTe, giving a metallic electronic structure. 
The Hubbard Hamiltonian term within the MnTe DFT+U calculation was parameterised with $U=4$~eV, $J=0.97$~eV (used previously for antiferromagnetic MnTe~\cite{Kriegner_PRB_2017}) and the fully-localised-limit (FLL) approximation to the double counting term.

\begin{figure*}[t!]
 \centerline{\includegraphics[width=\linewidth]{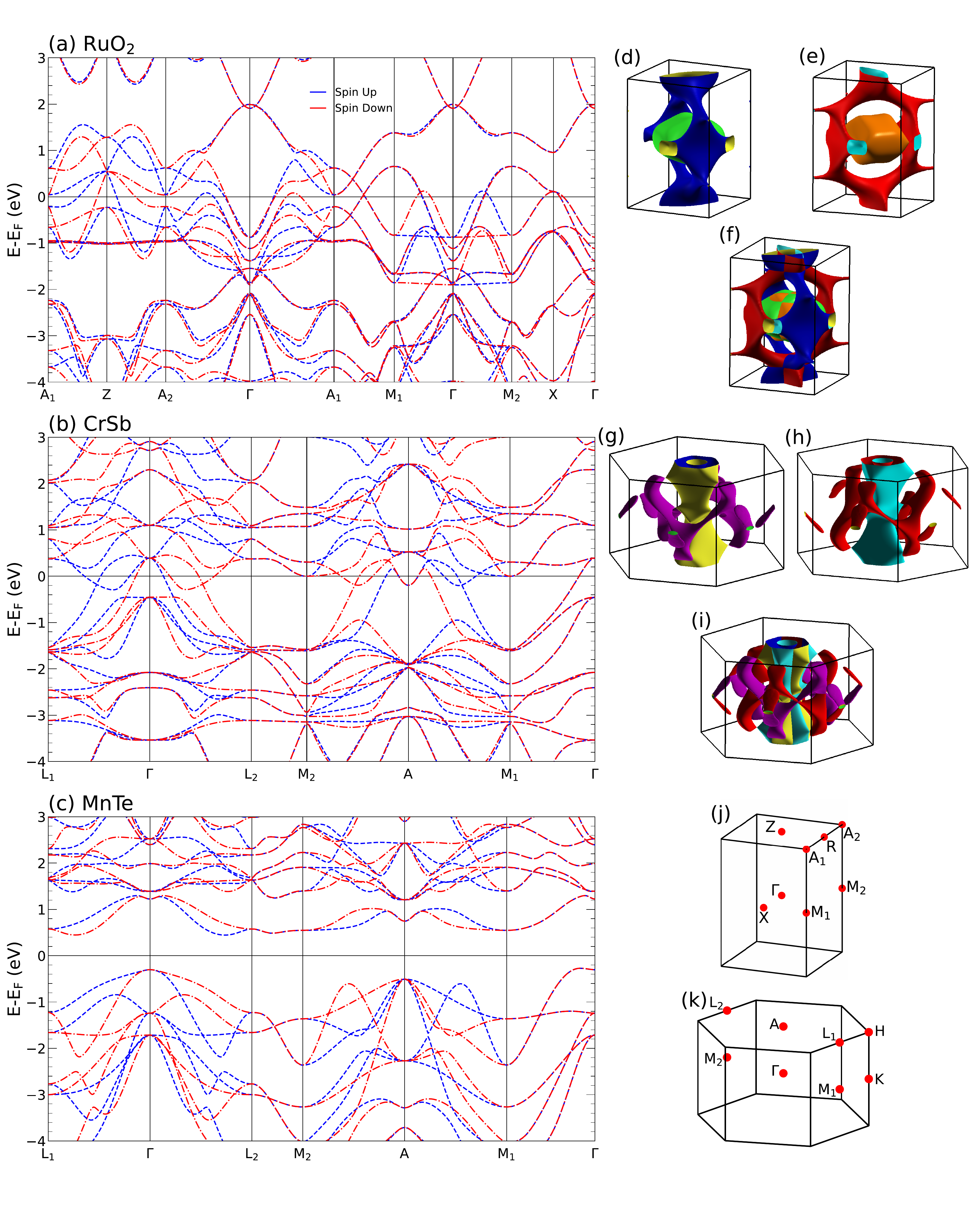}} 
 \caption{(a)-(c) The altermagnetic Band structures for RuO$_2$, CrSb and MnTe, respectfully. The high symmetry points are defined in the Brilluoin zones shown in (j)-(k). The chosen paths highlight the breaking of the spin degeneracy in the altermagnetic state of each material. For RuO$_2$, the spin-up and spin-down Fermi surface sheets are shown in (d) and (e), and all of the Fermi surface sheets are shown in (f). For CrSb, (g), (h) and (i) show the spin-up, spin-down and all the Fermi surface sheets, respectfully. The Fermi surface topology of each spin are the same as each other but rotated in the $k_x$-$k_y$ plane by $\pi/2$ radians for RuO$_2$ and by $\pi/3$ radians for CrSb.} 
\label{bands}

\end{figure*}

\begin{figure*}[t!]
 \centerline{\includegraphics[width=0.8\linewidth]{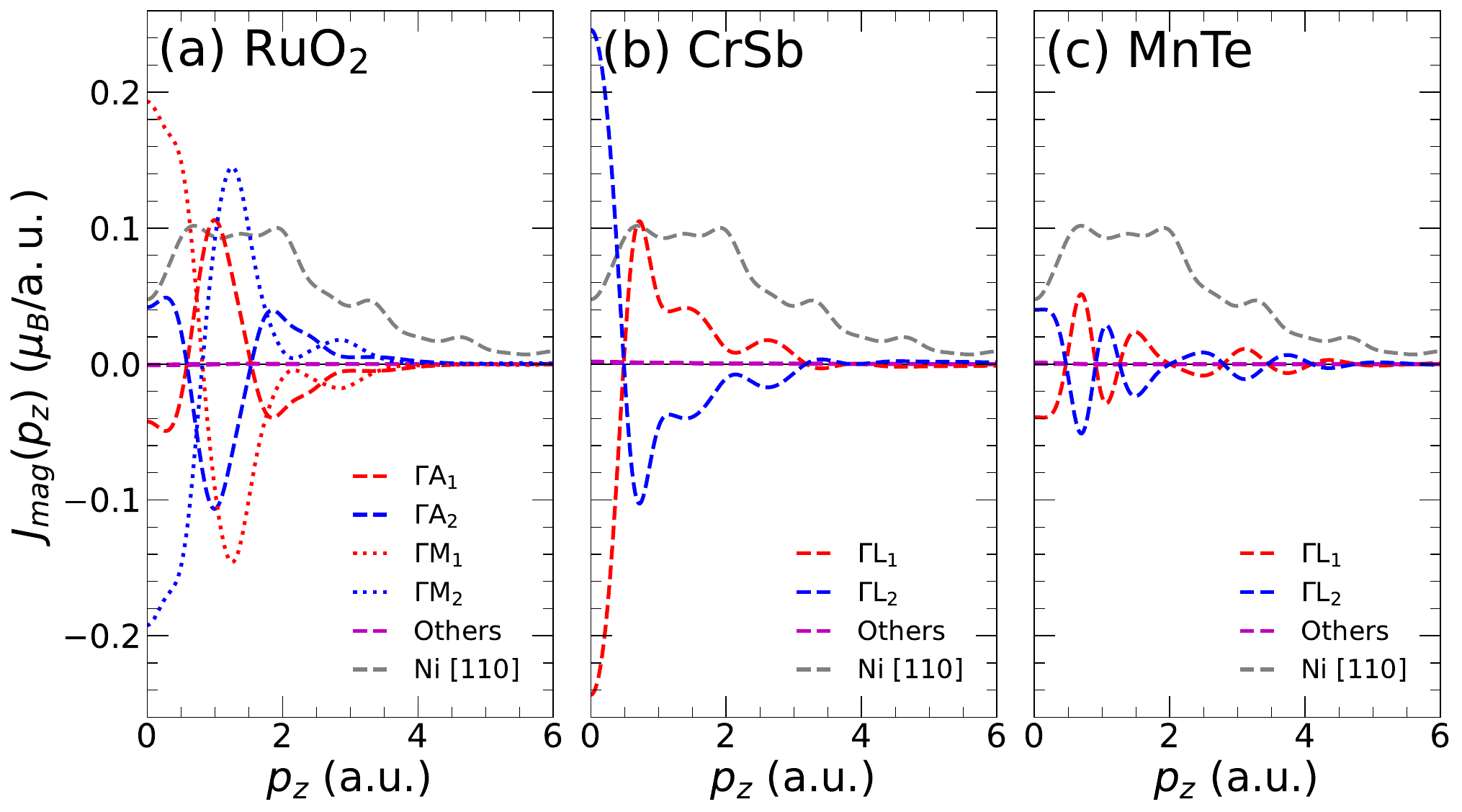}} 
 \caption{Predicted MCPs of altermagnetic (a) RuO$_2$, (b) CrSb and (c) MnTe. Each show the scattering vectors along which a significant MCP signal with respect to the shown Ni [011] MCP from Ref.~\cite{james2020magnetic}. The legend labels of these vectors (which pass through $\Gamma$) are based on the symmetry points given in Fig.~\ref{bands} (j) and (k). The altermagnetic MCPs of other directions along $\Gamma$ to a high symmetry point on the Brillouin zone edge results in $J_{\mathrm{mag}}(p_z)=0$ where the spins are degenerate. The inclusion of the Ni [011] MCP gives an indication of the expected experimental signal for these MCPs. We convolute our MCPs with the estimated experimental resolution.} 
\label{MCPs}

\end{figure*}

\begin{figure*}[t!]
 \centerline{\includegraphics[width=0.85\linewidth]{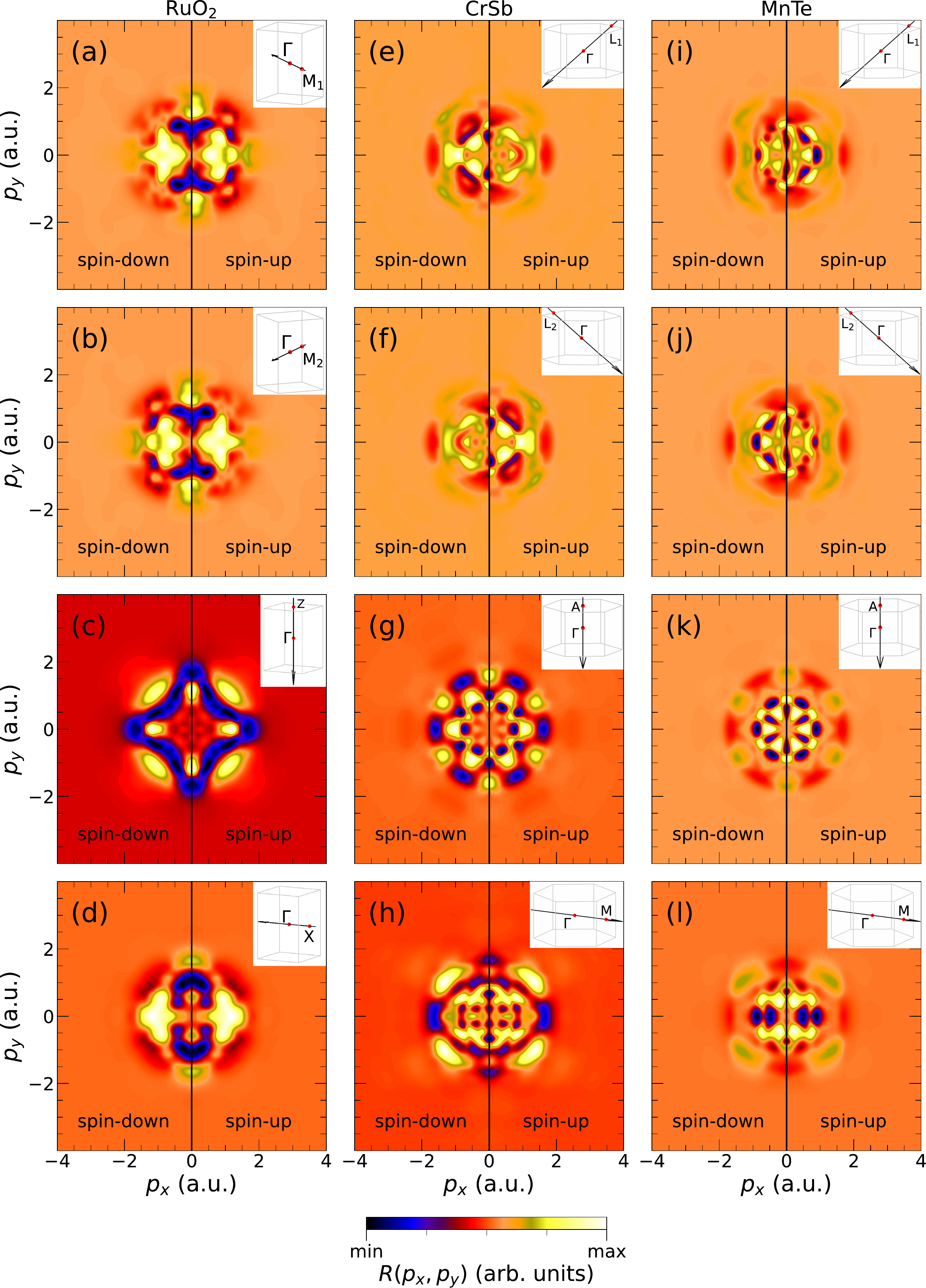}} 
 \caption{The 2D radial anisotropy of the 2D projected TPMDs along different projection vectors for (a)-(d) RuO$_2$, CrSb (e)-(h) and MnTe (i)-(l). The inset of each panel is a visual aid of the projection of the 3D TPMD along the vector (plane normal) going through the Brillouin zone. 
 The $p_x$ and $p_y$ axes are labelled as such for convention. The $p_x$ axes are parallel to the vector linking the high symmetry points: (a) -- $\Gamma$ to M$_{\rm 2}$; (b) -- M$_{\rm 1}$ to $\Gamma$; (c)-(d) -- M$_{\rm 1}$ to X; (e), (g)-(i) and (k)-(l) -- M$_{\rm 1}$ to K; and (f) and (j) -- K to M$_{\rm 1}$. The $p_y$ axes varies between the panels and are not always parallel to a high symmetry path.
 The first two rows of the panels present the projections which clearly show the differences in the spins where the last two panel rows show the spin degeneracy, all clearly highlighting key features in the altermagnetic state.} 
\label{Radial_aniso}

\end{figure*}

\begin{figure*}[t!]
 \centerline{\includegraphics[width=0.94\linewidth]{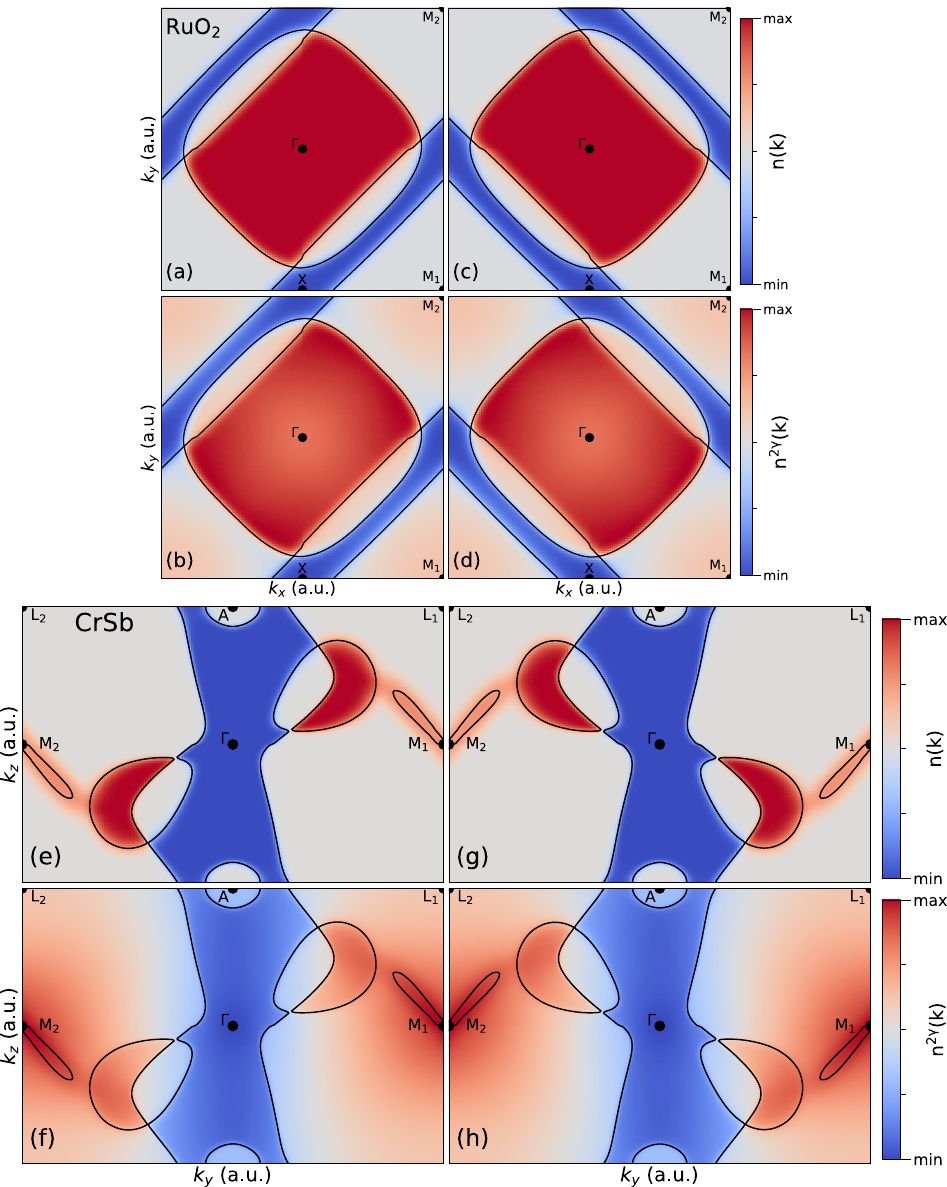}} 
 \caption{The 2D slices of the occupation distributions within the Brillouin zone of the metallic altermagnetics (a)-(d) RuO$_2$ and (e)-(h) CrSb. The left column shows the up spin distribution and the down spin is shown on the right column. Panels (a), (c) show the RuO$_2$ spin up and down electron occupation $n(k)$ slices in the k$_{\rm z} = 0$ plane. On the other hand, panels (b) and (d) show the spin up and down electron occupation as seen by the positron $n^{2\gamma}(k)$ in that plane. Panels (e) and (g) show the CrSb spin up and down $n(k)$ in the k$_{\rm x} = 0$ plane. Whereas panels (f) and (h) show the spin up and down $n^{2\gamma}(k)$ in that plane. For each panel, the Fermi surface contours from the energy eigenvalues are plotted for comparison. The Fermi surface are also located at the steps in the occupation distributions. These occupation distributions are not convoluted with an approximate experimental resolution function. The high symmetry points are labelled as per Fig.~\ref{bands}. } 
\label{Fig_4}

\end{figure*}

To calculate the EMDs and TPMDs and their projected quantities, we used the methodology implementation by Ernsting {\it et al.}~\cite{Ernsting_2014}. 
We used a maximum momentum cut-off of 8 a.u. for the EMDs and 4 a.u. for the TPMDs, by which momenta the densities have converged to zero. To simulate the measurements, we convoluted the 2D projected TPMD with a 2D Gaussian resolution function with a conservative estimate for the full-width-at-half-maximum of $\sim$ $1$ mrad ($0.137$ a.u.) to represent the resolution of the (spin-polarised) ACAR technique~\cite{ceeh_dissertation,Dugdale_2013}. For the MCPs, these were convoluted with a Gaussian resolution function with a full-width-at-half-maximum of $0.42$ a.u. which represents the momentum resolution of a typical magnetic Compton scattering experiment~\cite{Duffy_2013}. We normalise the total integral of the EMD to be equal to the number of valence electrons. With this normalization, the areas under the MCPs (their total integrals over momentum) are equal to the spin magnetic moment, which is zero for the altermagnets. For an indication of the experimental signal expected for the altermagnetic MCPs, the Ni [110] DFT MCP results from Ref.\cite{james2020magnetic} are also included. For direct comparison purposes, the Ni MCP was normalised such that the entire MCP area equals to the DFT spin moment (of 0.64 $\mu_{\rm B}$~\cite{james2020magnetic}), whereas the Ni experimental data was normalised such that the MCP for $\bf p \geq$~0 equals to the spin moment ~\cite{di.du.98} (so the theoretical MCP area is typically scaled by a factor of 2 for direct comparisons with this experimental data). 
The calculated spin momentum densities are linked to each other with the appropriate symmetry group~\cite{PhysRevX.12.031042}. However, very minor but visible discrepancies were present due to the symmetry breaking within the calculations in order to obtain the altermagnetic state which resulted in the spins being treated inequivalently (as the full set of symmetry operations were not found). Therefore, we symmetrised the spin momentum densities with respect to each other to remove these artificial discrepancies. This highlights how sensitive these techniques can be in showing the differences between the spin channels.

\section{Results}

\subsection{Band Structures and Fermi Surfaces}

For a long time, RuO$_2$ was been considered to be a Pauli paramagnet. However, recent neutron scattering experiments have shown that the two Ru atoms in RuO$_2$ each possess a magnetic moment of $\approx$ 0.05~$\mu_B$ 
which are aligned in an antiparallel fashion \cite{PhysRevLett.118.077201}, resulting in zero net magnetisation. Subsequent theoretical studies suggest that RuO$_2$ is actually an altermagnetic material~\cite{Kyo-Hoon_2019,PhysRevX.12.031042,doi:10.1126/sciadv.aaz8809}, as seen by the lifting of the spin degeneracy in the band structure and the Fermi surface sheets which break the time reversal symmetry. However, recent experiments~\cite{RuO2-QO-prx,kessler2024absence} contradict the interpretation that this material is indeed altermagnetic. Fig.~\ref{bands} shows our calculations of these quantities for RuO$_2$, where the band path in Fig.~\ref{bands} (a) was chosen to highlight paths where the breaking of the spin degeneracy is evident as well as some directions where the band energies of the spins are degenerate. The two Ru magnetic atoms in the RuO$_2$ unit cell have opposite spin but are connected by the $C_{4z}$ symmetry group which leads to a planar-like symmetry with $\pi/2$ rotations, protecting the net zero moment. The regions of the band path where the spin is not degenerate have an equivalent behaviour in the opposite spin in other regions of the Brillouin zone. This is observed, for example, in the band structure and Fermi surfaces in Fig.~\ref{bands} where the bands of one spin along M$_{\rm 1}$--$\Gamma$ are the same as the bands of the opposite spin along $\Gamma$--M$_{\rm 2}$. The presented band structure and Fermi surface are in agreement with those shown in Refs.~\cite{Kyo-Hoon_2019,PhysRevX.12.040501}. We note there are minor discrepancies which may arise from the certain parameters such as the muffin-tin radii or U and J value used, as has been observed in other calculations~\cite{PhysRevLett.118.077201, doi:10.1126/sciadv.aaz8809,Kyo-Hoon_2019}. These discrepancies do not significantly change the signatures of the altermagnetic state as seen in the radial anisotropies and MCPs presented later. The magnitude of the calculated ordered moment is 0.7~$\mu_B$ which is close to the one calculated in Refs.~\cite{Kyo-Hoon_2019,PhysRevLett.118.077201}, but both are notably different from the experimental ordered moment of approximately 0.05~$\mu_B$ at ambient conditions~\cite{PhysRevLett.118.077201}. However, when using $U$ and $J$ parameters which replicates the measured ordered moment, the splitting of the spin degeneracy is significantly suppressed meaning that the signatures of the altermagnetic state are unlikely to be measurable by spin-resolved ACAR or magnetic Compton scattering. However, we will continue with the discussion of the altermagnetic results with the parameters from Ref.~\cite{Kyo-Hoon_2019}.

CrSb was originally thought to be a traditional antiferromanget~\cite{snow1952neutron} but has recently been identified as an altermagnet from a theoretical perspective~\cite{PhysRevX.12.031042,guo2023spin}. This can be seen in the lifting of the spin degeneracy in the band structure and the Fermi surface sheets, as reproduced in our own CrSb calculations in Fig.~\ref{bands}. Recent experiments on CrSb thin films have observed band splittings and spin-integrated Fermi surfaces consistent with those predicted by DFT calculations in the altermagnetic state~\cite{reimers2024direct}. 
Our band structure and Fermi surface are in agreement with those shown in Refs.~\cite{PhysRevX.12.031042,guo2023spin}. The magnitude of the calculated ordered moment is 2.8~$\mu_B$ which is close to the experimental ordered moment of approximately 2.5~$\mu_B$ at ambient conditions~\cite{yuan_2020}. 

As with CrSb, MnTe was also thought to be a traditional antiferromagnet based on neutron scattering experiments~\cite{MnTe1965neutronAFM,MnTe-neutron-2005,Kriegner_PRB_2017}. However, inspired by studies of altermangetism described by spin groups~\cite{PhysRevX.12.040501}, MnTe has been proposed as a potential altermagnet~\cite{PhysRevX.12.040501,PhysRevX.12.031042}. Angle-resolved photoemission spectroscopy (ARPES) studies have revealed band splitting at low temperatures, suggesting a possible lifting of Kramers degeneracy induced by altermagnetic ordering~\cite{lee2024broken,krempasky2024altermagnetic}. Moreover, the presence of anomalous Hall effect (AHE)~\cite{PhysRevB.110.155201,PhysRevLett.130.036702} and the distinctive XMCD signal in the MnTe films~\cite{PhysRevLett.132.176701} indicate the breaking of time-reversal symmetry, thereby providing evidence for the existence of altermagnetic order. The magnitude of the calculated ordered moment is 4.4~$\mu_B$ which is close to the previous theoretical results~\cite{Kriegner_PRB_2017}, and the experimental ordered moment is approximately 4.7-5$\mu_B$~\cite{Kriegner_PRB_2017}. The altermagnetic state of insulating MnTe is clear to see in  Fig.~\ref{bands}. 
We note that we observe an altermagnetic band structure in our metallic MnTe DFT results, which also leads to significant signatures of the altermagnetic state in its MCPs and TPMD. 

The $\mathbf{k}$-path of CrSb and MnTe in Fig.~\ref{bands} (b) and (c) was chosen to highlight the breaking of the spin degeneracy along the paths between high symmetry points in the irreducible Brillouin zone (see Fig.~\ref{bands} (k)). The two Cr/Mn atoms with opposite magnetic moments in the crystal structure are connected by the $C_{6z}$ symmetry group which results in the unusual bulk-like symmetry of spin textures with a $\pi/3$ rotation in reciprocal space and again preserving the zero net magnetic moment~\cite{PhysRevX.12.031042}. The consequence of this can be seen in the band structure and Fermi surfaces in Fig.~\ref{bands} where, for example, the bands of one spin along L$_{\rm 1}$--$\Gamma$ are the same as the bands of the opposite spin along $\Gamma$--L$_{\rm 2}$. 

For the metallic altermagnets, slices of the Fermi surface sheets shown in Fig.~\ref{bands} (f) and (i) would be experimentally observed by several techniques such as ARPES. For ARPES measurements of RuO$_2$, Ref.~\cite{PhysRevB.98.241101,OF-RuO2-ARPES-AM} only distinguished the time reversal breaking but not clearly distinguish the contributions from each spin. Spin-resolved ARPES experiments have also been applied to study the altermagnetism in CrSb~\cite{yang2025three}. By extracting spin-resolved quasiparticle spectra along one-dimensional cuts in selected regions of momentum space, Yang et al.~\cite{yang2025three} were able to assign the spin-integrated bands to their respective spin components. However, a complete spin-resolved Fermi-surface reconstruction was not achieved. 
All of the Fermi surface sheets would be experimentally observed by other techniques such as Quantum Oscillations and both {\it spin-integrated} Compton scattering and 2D-ACAR. However, spin polarised 2D-ACAR would be able to \textit{clearly} resolve these spin contributions. 
For both RuO$_2$ and CrSb, there are two Fermi surface sheets for each spin which are shown in Figs.~\ref{bands} (d)-(e) and (g)-(h). The topologies of the Fermi surface for each spin are the same with the only difference being a rotation in the $k_x$-$k_y$ plane connected by the aforementioned corresponding symmetry group.

\subsection{Magnetic Compton Profiles}

The MCPs for the altermagnets are presented in Fig.~\ref{MCPs} which highlights the altermagnetic state, following that presented for MnF$_2$ by Bhowal and Spaldin \cite{PhysRevX.14.011019}. As expected, our MCPs are antisymmetric along the scattering vectors where the spin is non-degenerate. However, in contrast to their predictions for MnF$_2$ MCPs, the magnitude of our antisymmetric MCPs are very comparable in magnitude to that predicted for ferromagnetic Ni which has been measured previously by Dixon $\textit{et al.}$~\cite{di.du.98}. Along the other $\Gamma$ to high symmetry point directions, the MCP is equal to zero over the entire momentum range as expected for spin degenerate directions. As expected, the integral of the MCPs, which is the spin moment, is equal to zero (within numerical noise) for each direction. Even the sum of the pair of corresponding altermagnetic MCPs, where these MCPs have significant signal, gives zero spin moment. Clearly, the spin-polarised EMD can vary throughout momentum space, as long as its volume integral is equal to zero to conserve the zero net moment. There are fewer scattering vectors which display the altermagnetic state within the MCP than the altermagnetic signatures along the $k$-paths shown in Fig.~\ref{bands}. This is due to the projection nature of the EMDs to get the MCP, where the altermagnetic spin non-degenerate regions are integrated away along certain scattering vectors. However, each material are strong candidates for magnetic Compton scattering measurements to show signatures of the altermagnetic state.

\subsection{Electron-Positron Radial Anisotropies}

Clear signatures of the altermagnetic state are visible within the predicted radial anisotropies of the 2D projected spin-resolved TPMDs as shown in Fig.~\ref{Radial_aniso} for each material irrespective whether it is metallic or insulating. For CrSb and RuO$_2$, these difference are partly owing to the different projected Fermi surface topologies of the spin-up and spin-down electrons. However, it is interesting that there is still a strong anisotropy in the insulating MnTe. Just like with the MCPs, these radial anisotropy signals are strong and experimentally discernable. 
The radial anisotropies from the projected TPMD along the directions specified within the panel insets of the first two rows of Fig.~\ref{Radial_aniso} are spin non-degenerate with clear distinct differences between each spin. Again, this is due to the projection vectors being along the spin non-degenerate directions where the altermagnetic signatures are not integrated away. As expected, the `up' spin radial anisotropies (right side of each panel) of the first row of Fig.~\ref{Radial_aniso} are equivalent to the `down' spin radial anisotropies (left side of each panel) of the second row of Fig.~\ref{Radial_aniso}. The same is true for the `down' spin radial anisotropies of the first row of Fig.~\ref{Radial_aniso} being equivalent to the `up' spin radial anisotropies of the second row of Fig.~\ref{Radial_aniso}. The other important signature of the altermagnetic state is the absence of any differences between the radial anisotropies of each spin along other projected directions, such as those shown in the last two rows of Fig.~\ref{Radial_aniso}. Seeing these significant differences, and the absence of any difference between radial anisotropies along particular projection vectors, the spin-resolved radial anisotropies from measured 2D-ACAR spectra would be be a strong affirmation of the altermagnetic state within metallic and insulating materials. 

\subsection{Occupation Distribution}

A full spin-resolved Fermi surface reconstructed from a set of spin-resolved momentum densities obtained from a set of spin-polarised 2D-ACAR spectra would also provide unequivocal proof of the presence of altermagnetism in metals. The spin-resolved 3D Fermi surfaces can be reconstructed from a set of spin-polarised 2D-ACAR spectra as described by Weber {\it et al.} \cite{PhysRevLett.115.206404}. The set of separated spin-resolved 2D TPMDs, as measured from the ACAR spectra, from which the 3D TPMD and thus the spin-resolved Fermi surfaces would be reconstructed. These calculated Fermi surfaces show that the topology would be distinguishable with the typical precision of ACAR. To illustrate this reconstruction for the metallic altermagnetics, Fig.~\ref{Fig_4} shows 2D slices of both the electron occupation n$(\bf \rm k)$ and the electron occupation as seen by the positron n$^{\rm 2\gamma}(\bf \rm k)$. The spin-resolved n$(\bf \rm k)$ shows the contrast to the experimentally determinable n$^{\rm 2\gamma}(\bf \rm k)$ distributions, which is what would be determined from the spin-resolved reconstruction. Even convoluting the n$^{\rm 2\gamma}(\bf \rm k)$ distributions in Fig.~\ref{Fig_4} with a resolution function representative of the measurement and reconstruction gives the Fermi surface features which are significantly and robustly different between each spin. Of course, these occupation distributions for both spins would be equivalent after applying the appropriate symmetry. The Fermi surface would be determined from the significant occupation steps in the distribution. In Fig.~\ref{Fig_4}, the Fermi surface features are still present within the  n$^{\rm 2\gamma}(\bf \rm k)$ distributions when comparing to the corresponding n$(\bf \rm k)$. However, the positron does have a pronounced effect on the n$^{\rm 2\gamma}(\bf \rm k)$ distribution. For RuO$_2$, the influence of the positron enhances and de-enhances certain regions of the n$^{\rm 2\gamma}(\bf \rm k)$ distribution. This is to be expected, and has been seen previously~\cite{Kaiser_1986,dugdale_2014,Dugdale_2013}. The extent of the enhancement from the positron manifests itself within the electron-positron matrix elements, and as such depends on the nature of the orbital character of the bands, which has been discussed previously~\cite{PhysRevB.82.125127}. 
This is akin to the matrix elements seen within ARPES which can suppress spectral features in the measurements~\cite{MOSER201729}. These APRES matrix elements would need to be taken into account for comparisons between the ARPES data of altermagnetic candidates and their predicted band structures, although direct observation of distinctive band structure features were still possible without their inclusion~\cite{reimers2024direct,OF-RuO2-ARPES-AM}. This situation is somewhat similar for these theoretical positron-related results, where the n$^{\rm 2\gamma}(\bf \rm k)$ and radial anisotropies have clear and distinct features signifying the altermagnetic state. 
We note that the electron-positron enhancement $\gamma(\mathbf{r})$ also contributes to the occupation enhancement too, but the influence of this is not as significant as the electron-positron matrix elements. 
For CrSb, on the other hand, the positron has notably enhanced the distribution around the M symmetry points and the nearby L symmetry points. There is a Fermi surface sheet around M which comes from the band just grazing the Fermi level in Fig.~\ref{bands} (b). Such a band would be difficult to measure in other experimental techniques including ARPES and Quantum Oscillations. 
Nonetheless, The spin-resolved Fermi surface features are still distinguishable within the electron-positron occupation despite these positron enhancements. Therefore, these distributions for both materials will provide an unequivocal distinction of the spin-dependent Fermi surface features, especially when compared to techniques measuring both spins, such as 2D-ACAR, Compton scattering, ARPES and Quantum Oscillations.

\section{Conclusion}

We propose that momentum density  spectroscopy, such as spin-polarized 2D-ACAR and magnetic Compton scattering, could be powerful probes of altermagnetism in metallic and insulating candidates. Using RuO$_2$, CrSb and MnTe as examples, we show how the peculiar magnetic order leads to specific and distinct signatures which should be visible in spin-resolved 2D-ACAR spectra through their radial anisotropies and in the magnetic Compton profiles from magnetic Compton scattering measurements. We note that our altermagnetic magnetic Compton profiles show similar oscillatory structure to those calculated by Bhowal and Spaldin~\cite{PhysRevX.14.011019}, but the magnitudes of our calculated magnetic Compton profiles are comparable to the calculated Ni ones which indicate that these are indeed measurable. 
We also suggest that a full 3D reconstruction of the Fermi surfaces of each spin would be feasible from a set of spin-polarised 2D-ACAR spectra and would also provide unequivocal proof of the presence of altermagnetism. This unambigious extraction of each altermagnetic spin Fermi surface by spin-polarised 2D-ACAR is difficult by more commonly used electronic and magnetic structure probes. As described by Weber {\it et al.}~\cite{PhysRevLett.115.206404}, a set of separated spin-resolved 2D TPMDs as measured from the 2D-ACAR spectra would be extractable, from which the 3D TPMD and so the spin-resolved Fermi surfaces would be reconstructed. Extracting and analysing the altermagnetic spin-resolved Fermi surfaces from spin-resolved positron annihilation experiments would be key to unravel intriguing physical effects of the altermagnets and their potential applications. From these results, we strongly recommend altermagnetic candidates to be measured by both spin-polarised ACAR and magnetic Compton scattering.

\section{Acknowledgements}
Wenhan Chen acknowledges the funding and support from the Chinese Scholarship Council (CSC), Grant No. 201908060087. Calculations were performed using the computational facilities of the Advanced Computing Research Centre, University of Bristol (\href{http://bris.ac.uk/acrc/}{http://bris.ac.uk/acrc/}). XCrysDen~\cite{kokalj_1999} has been used in the preparation of some figures.

\bibliography{ref}

\end{document}